\documentclass[twocolumn,floats,showpacs,prl,amssymb,floatfix,nofootinbib,balancelastpage,superscriptaddress,amsmath]{revtex4}
\usepackage{epsfig}
\def\sss{\scriptscriptstyle}

                              \newlength{\strikewidth}
                              \newlength{\strikelength}
\begin{document}

\title{Non-thermal X-rays from the Ophiuchus galaxy cluster and dark matter annihilation}

\author{Stefano Profumo}
\email{profumo@scipp.ucsc.edu}
\affiliation{%
Santa Cruz Institute for Particle Physics and Department of Physics,\\ University of California, Santa Cruz CA 95064
}%

\begin{abstract}
\noindent  We investigate a scenario where the recently discovered non-thermal hard X-ray emission from the Ophiuchus cluster originates from inverse Compton scattering of energetic electrons and positrons produced in weakly  interacting dark matter pair annihilations. We show that this scenario can account for both the X-ray and the radio emission, provided the average magnetic field is of the order of \mbox{0.1 $\mu$G}. We demonstrate that GLAST will conclusively test the dark matter annihilation hypothesis. Depending on the particle dark matter model, GLAST might even detect the monochromatic line produced by dark matter pair annihilation into two photons.

\end{abstract}




\maketitle

Clusters of galaxies are the largest bound dark matter (DM) structures in the universe. As such, they are natural targets for the search for observational signatures of particle DM \cite{Colafrancesco:2005ji}. If DM is in the form of weakly interacting massive particles (WIMPs) \cite{WIMPReviews,KKDM}, DM pair annihilations generically produce $\gamma$-rays as well as a non-thermal energetic electron-positron ($e^\pm$) population. The latter, in turn, is expected to yield secondary emissions at soft $\gamma$-ray, X-ray and radio frequencies via inverse Compton scattering, bremsstrahlung and synchrotron radiation, opening up the possibility of a multi-wavelength approach to particle DM detection \cite{Colafrancesco:2005ji,Colafrancesco:2006he}. A generic feature of the broad-band DM annihilation spectrum is a significant hard X-ray component \cite{Colafrancesco:2005ji,xrgrinprep}.

Interestingly, the discovery of a non-thermal hard X-ray emission from the Ophiuchus cluster, detected with relatively robust statistical significance in a 3 Ms observation with the IBIS/ISGRI and JEM-X instruments on board INTEGRAL, was recently reported in Ref.~\cite{Eckert:2007fv}.  The Ophiuchus cluster is a nearby ($z\simeq0.028$ \cite{johnston}) rich cluster with a high temperature plasma ($kT\sim10$ keV), featuring the second brightest emission in the 2-10 keV band. In addition to X-ray observations, the steep-spectrum radio source MSH 17-203 \cite{radiodata} was associated to the Ophiuchus cluster \cite{johnston}, indicating the presence of relativistic electrons. However, the cluster was not detected at $\gamma$-ray frequencies by EGRET \cite{Reimer:2003er}.

Several clusters are known to host extended radio emissions \cite{feretti}, suggesting the existence of energetic non-thermal electrons that radiate at radio frequencies through synchrotron emission. The same electron population should also produce hard X-rays via inverse-Compton (IC) scattering off cosmic microwave background photons. Up to now, however, firm evidence for extended non-thermal hard X-ray emission in clusters was still missing, with a few controversial \cite{Rossetti:2003vy,Pfrommer:2007sm} exceptions, including BeppoSAX \cite{FuscoFemiano:1999em} and INTEGRAL \cite{Renaud:2006ch} observations of the Coma cluster, BeppoSAX observations of Abell 2256 \cite{FuscoFemiano:2005tz} and Chandra observations of the Perseus cluster \cite{Sanders:2005jx}. In addition, under the assumption of negligible contamination from obscured AGNs, \cite{neva} reports $\sim2\sigma$ detections, using the BeppoSAX PDS instrument, of hard X-ray non-thermal components from a few more merging clusters, including Abell 2142, 2199, 3376, Virgo and the Ophiuchus cluster itself. 

While ordinary astrophysical mechanisms, including merger shocks, can be invoked to explain the non-thermal electrons presumably responsible for the observed non-thermal activity in galaxy clusters, in the present analysis we propose and investigate a novel scenario where WIMP annihilations produce, or significantly contribute to, said non-thermal population responsible for the hard X-ray detection in the Ophiuchus galaxy cluster. In a model independent approach, we determine the parameters of the particle DM setups that provide the best fits to the INTEGRAL X-ray data, and we compute the resulting multi-wavelength spectra. We then compare these spectra with the radio data and with the $\gamma$-ray limits and future prospects for the soon-to-be-launched Gamma-Ray Large Area Telescope (GLAST). The highlights of our analysis are: (1) the DM hypothesis will conclusively be probed with GLAST; (2) the radio emission can in principle also be fitted with the synchrotron emission from DM-annihilation-produced $e^\pm$, as long as the average magnetic field in the cluster is of the order of 0.1 $\mu$G; (3) GLAST might be able detect the monochromatic $\gamma$-rays produced in direct DM pair annihilation into two photons.

The flux of $e^\pm$ produced by WIMP pair annihilations depends on the particle DM setup and on the DM density distribution. We define a source function $Q_e(E_e,\vec x)$, which gives the number of $e^\pm$ per unit time, energy and volume element produced locally in space, as
\begin{equation}
Q_e(E_e,\vec x)=\langle\sigma v\rangle_0\sum_f\frac{{\rm d}N^f_e}{{\rm d}E_e}(E_e)\ B_f\ {\cal N}_{\rm pairs}(\vec x).
\end{equation}
In the equation above, $\langle\sigma v\rangle_0$ is the WIMP annihilation rate at zero temperature, the sum is over all kinematically allowed Standard Model annihilation final states $f$, each with a branching ratio $B_f$ and an $e^\pm$ distribution ${\rm d}N_e^f/{\rm d}E_e$, and  ${\cal N}_{\rm pairs}(\vec x)$ is the number density of WIMP pairs at a given point $\vec x$, i.e. the number of WIMP particle pairs per volume element squared: ${\cal N}_{\rm pairs}(\vec x)=\rho_{\rm\sss DM}^2(\vec x)/(2m_{\rm\sss DM}^2)$, where $\rho_{\rm\sss DM}$ stands for the DM density. The particle physics framework sets the quantity $\langle\sigma v\rangle_0$, the list of $B_f$ and the mass of the WIMP, $m_{\rm\sss DM}$. The latter also determines the energy scale of the pair annihilation event, and, together with the specific final state $f$, the ${\rm d}N_e^f/{\rm d}E_e$ spectral functions, which we numerically compute with the Monte Carlo code Pythia \cite{pythia}. In addition, $m_{\rm\sss DM}$ enters in the determination of the local number density of WIMP pairs. 

Once the source function $Q_e(E_e,\vec x)$ is determined, the $e^\pm$ spectrum and density are affected by spatial diffusion and energy loss processes, usually described -- under the assumptions of negligible convection and re-acceleration effects -- by a diffusion-loss equation of the form 
\begin{eqnarray}\label{eq:dl}
\frac{\partial}{\partial t}\frac{{\rm d}n_e}{{\rm d}E_e}(E_e,\vec x)=\vec\nabla\cdot\Big[D(E_e,\vec x)\vec\nabla\frac{{\rm d}n_e(E_e,\vec x)}{{\rm d}E_e}\Big]\nonumber &&\\
+\frac{\partial}{\partial E_e}\Big[b(E_e,\vec x)\frac{{\rm d}n_e(E_e,\vec x)}{{\rm d}E_e}\Big]+Q_e(E_e,\vec x),&&
\end{eqnarray}
where ${\rm d}n_e/{\rm d}E_e$ is the number density of electrons per unit energy, $D$ is the diffusion coefficient, and
\begin{eqnarray}\label{eq:enloss}
 b(E_e,\vec x) &= &  b_{\rm IC} + b_{\rm syn} + b_{\rm Coul} + b_{\rm brem}
 \end{eqnarray}
encodes the various energy loss mechanisms \cite{Colafrancesco:2005ji}. 

Knowledge of the distribution of the DM-induced $e^\pm$ population ${\rm d}n_e/{\rm d}E_e$, of the magnetic field structure and strength, as well as of the electron, gas and starlight densities allows one to compute the WIMP-induced secondary emissions. Specifically, at radio frequencies the DM-induced emission is dominated by the synchrotron radiation of the relativistic secondary electrons and positrons. IC scattering of the non-thermal $e^\pm$ on target CMB and starlight photons gives rise to a spectrum of photons stretching from below the extreme ultra-violet up to the soft $\gamma$-ray band, peaking in the X-ray energy band. Non-thermal bremsstrahlung, {\em i.e.} the emission of $\gamma$-ray photons in the deflection of the charged particles by the electrostatic potential of ionized gas, contributes in the soft $\gamma$-ray band. Finally, a hard $\gamma$-ray component arises from prompt emission in WIMP pair annihilations, mostly originating from the two photon decay of neutral pions, and, at the high energy end of the spectrum, from internal bremsstrahlung from charged particle final states. The $\gamma$-ray spectrum extends up to energies equal to the kinematic limit set by the WIMP mass, $E_\gamma\le m_{\rm\sss DM}$, and might feature one or more monochromatic lines associated to two-body annihilation final states where one (or both) of the particles is a photon. We refer the reader to Ref.~\cite{Colafrancesco:2005ji} for details on the computation of the multi-wavelength emission from DM annihilation.

\begin{figure}
\centerline{\epsfig{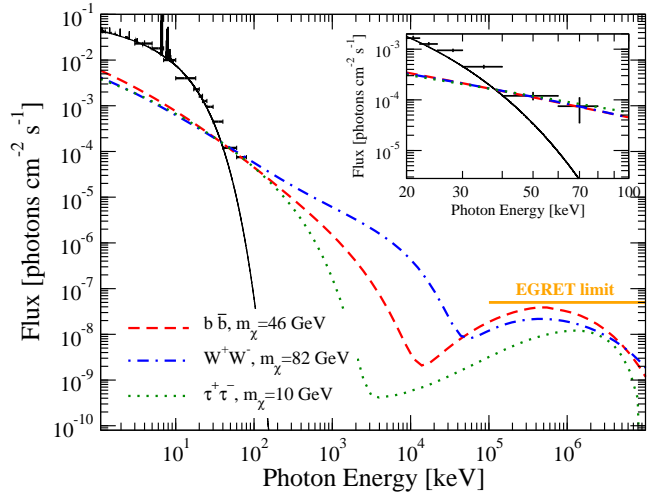}}
\caption{The hard X-ray and $\gamma$-ray spectrum for three DM particle models, plus a single-temperature MEKAL model \cite{kaastra} for the thermal X-ray emission, compared with the INTEGRAL data \cite{Eckert:2007fv} and with the EGRET upper limit \cite{Reimer:2003er}.}
\label{fig:spectrum}
\end{figure}
We show in Fig.~\ref{fig:spectrum} the photon flux in the hard X-ray and $\gamma$-ray bands for three benchmark DM particle model accounting for the INTEGRAL data. We also include a thermal component from the bremsstrahlung emission of the intra-cluster medium, obtained with a single-temperature MEKAL model \cite{kaastra} with the abundance fixed to 0.49 compared to the solar value and $kT=8.5$ keV \cite{Eckert:2007fv}. The thermal component was normalized to produce the best fit for the INTEGRAL data below 20 keV. The DM models were instead normalized to obtain the best global fit to the data above 20 keV. We chose DM models with $B_f=1$ for $f=b\bar b,\ W^+W^-,\ \tau^+\tau^-$, {\em i.e.} each model pair annihilating into a single Standard Model final state. The three particular final states were selected for two reasons: (1) the resulting $e^\pm$ spectra ${\rm d}N_e^f/{\rm d}E_e$ range from the softest ($b\bar b$) to the hardest ($\tau^+\tau^-$) possible case \cite{Colafrancesco:2005ji}; (2) the three final states correspond to common well-defined cases found in supersymmetric DM models \cite{WIMPReviews}. For instance, in the minimal supergravity scenario \cite{msugra} $f\simeq b\bar b$ corresponds to the so-called bulk and funnel regions where the neutralino has a relic abundance compatible with the cold DM density, $\tau^+\tau^-$ is found in the coannihilation region (where neutralino pair-annihilation proceeds predominantly through scalar tau exchange) and $W^+W^-$ in the focus point region \cite{focus}. This choice of benchmark models follows here and generalizes the approach of \cite{Colafrancesco:2005ji} -- linear combinations of the considered models produce almost any WIMP multi-wavelength emission spectrum.

For each final state, we selected the DM particle mass giving the lowest $\chi^2$ in the fit to the INTEGRAL data: $m_{DM}(f=b\bar b)=46$ GeV, $m_{DM}(f=W^+W^-)=82$ GeV and $m_{DM}(f=\tau^+\tau^-)=10$ GeV. Following \cite{refsergio} we assumed, for the diffusion coefficient, the form
\begin{equation}
\nonumber D(E_e)=D_0\frac{d_B^{2/3}}{B_\mu^{1/3}}\left(\frac{E_e}{1\ {\rm GeV}}\right),\quad D_0=3.1\times 10^{28}\ {\rm cm}^2{\rm s}^{-1},
\end{equation}
where $d_B\simeq20$ is the minimum scale of uniformity of the magnetic field in kpc and $B_\mu$ is the average magnetic field in $\mu$G. Notice that, while we take into account spatial diffusion in the numerical computation, the effect of changing parameters such as $D_0$ or $d_B$ is minimal on the resulting WIMP multi-wavelength spectrum. In the computation of the photon flux we neglected the IC from starlight and assumed an average thermal gas density $n^{\rm th}=10^{-3}\ {\rm cm}^3$, relevant for the computation of $b_{\rm Coul}$ and $b_{\rm brem}$ in Eq.~(\ref{eq:enloss}). In the portion of the spectrum shown in Fig.~\ref{fig:spectrum} the value of the magnetic field is not crucial, and was fixed here, for reference, to $B=0.1\ \mu$G. As shown in the inset, the fit to the INTEGRAL data improves dramatically with the contribution from DM annihilation. Also, the emission in the $\gamma$-ray band is compatible with the EGRET limit \cite{Reimer:2003er}, shown for reference with a horizontal orange line.

\begin{figure}
\centerline{\epsfig{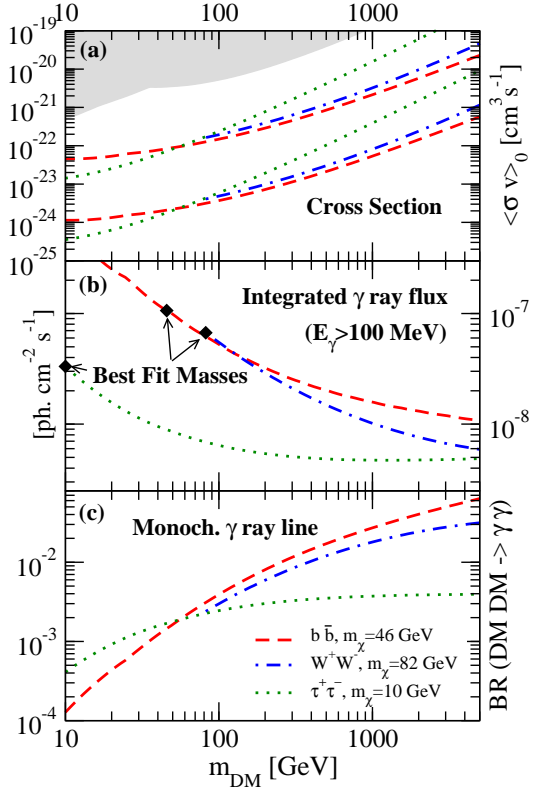}}
\caption{The preferred pair annihilation cross section (a), the integrated $\gamma$-ray flux above 100 MeV (b) and the minimal branching ratio for the detection of the monochromatic $\gamma$-ray line at $E_\gamma=m_{\rm DM}$ (c), as a function of the DM particle mass. In panel (a) the upper lines refer to the case of no substructures, the lower lines refer to the substructure setup described in the text, and the gray shaded region is ruled out by EGRET \cite{MayerHasselwander:1998hg} and H.E.S.S. \cite{Aharonian:2006wh} data on the gamma-ray flux from the galactic center region.}
\label{fig:gammarays}
\end{figure}
What is the DM pair annihilation cross section required to reproduce the spectra shown in Fig.~\ref{fig:spectrum}? To answer this question we need to integrate the number density of DM pairs over the line of sight. In turn, this requires knowledge of the DM density profile for the Ophiuchus cluster. We follow here the analysis of Ref.~\cite{Colafrancesco:2005ji}, and assume, for reference, the DM density profile obtained in the numerical simulations of Ref.~\cite{N04} (namely $\rho(r)=\rho_0 g(r/a)$, with $g(x)=\exp[-2(x^\alpha-1)/\alpha]$ and $\alpha\simeq0.17$) and the DM substructure setup outlined in \cite{bullock}. We verified that using other DM profiles changes our predictions by less than one order of magnitude. We derived from Ref.~\cite{reiprich} a virial mass of $\sim1.5\times 10^{15} M_\odot h^{-1}$ and a virial concentration of $\sim10$. We parametrize the contribution of substructures via the fraction $f_s$ of total mass in subhalos, and assume a ratio $R_s=5$ between the concentration parameter in subhalos and that in isolated halos with equal mass, $R_s\equiv\langle c_s\rangle/\langle c_{\rm vir}\rangle$. In Fig.~\ref{fig:gammarays} (a) we compute the cross section as a function of the DM particle mass, for the three benchmark final states, giving the best fit to the INTEGRAL data. We assume $f_s=0.5$ for the lower lines, as suggested by numerical simulations \cite{diemand}, while we neglect the contribution from substructures for the upper lines. 

While we find rather large values for the pair annihilation cross section compared to the naive expectation $\langle\sigma v\rangle_0\simeq 3\times 10^{-26}\ {\rm cm}^3{\rm s}^{-1}$ motivated by requiring a thermal WIMP relic abundance compatible with the CDM density through simple scaling arguments, the range we get is consistent with several examples of supersymmetric DM models (see {\em e.g.} Fig.~15 in \cite{Colafrancesco:2005ji}). Also, the values we obtain for both $\langle\sigma v\rangle_0$ and $m_{\rm\sss DM}$ are consistent with all available particle physics constraints on DM, and are compatible with WIMPs being in the right density today provided, for instance, non-thermal production or a modified cosmological expansion rate is assumed at the time of WIMP freeze-out \cite{nonthermal}. In addition, uncertainties on (1) the dark matter density distribution and (2) the galactic and extra-galactic gamma-ray background undermine the possibility of ruling out WIMP models with large pair-annihilation cross sections via gamma-ray data from the center of the Galaxy \cite{MayerHasselwander:1998hg, Aharonian:2006wh,Cesarini:2003nr,Profumo:2005xd}, nearby galaxies \cite{Evans:2003sc,Bergstrom:2005qk,Profumo:2006hs}, and from the galactic halo \cite{Moskalenko:2006zy}. For instance, in the upper panel of fig.~\ref{fig:gammarays} we shade in gray the region ruled out by EGRET and H.E.S.S. data from the galactic center region, assuming a cored dark matter profile following the analysis of Ref.~\cite{Cesarini:2003nr,Profumo:2005xd}. Clearly, gamma-ray data from the galactic center do not rule out the range of cross sections we find.

Similarly, antimatter constraints not only depend on the dark matter density across the hole galactic halo, but are also affected by sizable uncertainties in the diffusion and propagation of cosmic rays in the Galaxy \cite{Moskalenko:2006zy}. Even for cross sections as large as to account for the EGRET data on the galactic gamma-ray emission \cite{deBoer:2005tm} (hence at the level of the cross sections in the shaded gray region in fig.~\ref{fig:gammarays}), antimatter fluxes from WIMP annihilations are in general compatible with available data \cite{Gebauer:2007mg}.  Finally, both the search for energetic neutrinos from WIMP annihilation in the Sun or the Earth, and direct WIMP detection, depend on the WIMP-nucleon scattering cross section, a quantity which is unrelated to the WIMP pair annihilation. Even in special models, such as supersymmetry, the range of predictions is so wide to make it impossible to constrain $\langle\sigma v\rangle_0$ with neutrino fluxes or direct detection searches \cite{Kamionkowski:1994dp,Profumo:2004at}. In conclusion, the WIMP pair annihilation rates we consider here, while larger than what naively expected, are compatible with WIMP cosmological DM production and DM searches.

Panel (b) in  Fig.~\ref{fig:gammarays} shows the integrated $\gamma$-ray flux above 0.1 GeV for the best fit models as a function of $m_{\rm\sss DM}$. In all cases we find that the expected $\gamma$-ray flux is well above the anticipated GLAST LAT integral flux sensitivity, estimated to be around a few $\times 10^{-10}\ {\rm cm}^{-2} {\rm s}^{-1}$ \cite{latperform}. Fig.~\ref{fig:gammarays} (c) shows the branching ratio $\langle\sigma v\rangle_{\rm tot}/\langle\sigma v\rangle_{\gamma\gamma}$ for the monochromatic ${\rm DM\ DM}\rightarrow\gamma\gamma$ channel needed to obtain, for the best fit models, the detection of at least 10 photons\footnote{The branching ratio corresponding to a larger photon flux can be obtained by linearly rescaling the lines in Fig.~\ref{fig:gammarays} (c).} with $E_\gamma=m_{\rm\sss DM}$. Notice that the values shown are independent of the assumed DM profile. While $\langle\sigma v\rangle_{\rm tot}/\langle\sigma v\rangle_{\gamma\gamma}$ is entirely model dependent, the range we obtain is generically consistent with what is expected {\em e.g.} in supersymmetry \cite{ullioline}, and especially in next-to-minimal supersymmetric extensions of the Standard Model \cite{Ferrer:2006hy}. GLAST can therefore easily detect a sizable number of monochromatic energetic $\gamma$ rays, depending on the specific DM particle model.

\begin{figure}
\centerline{\epsfig{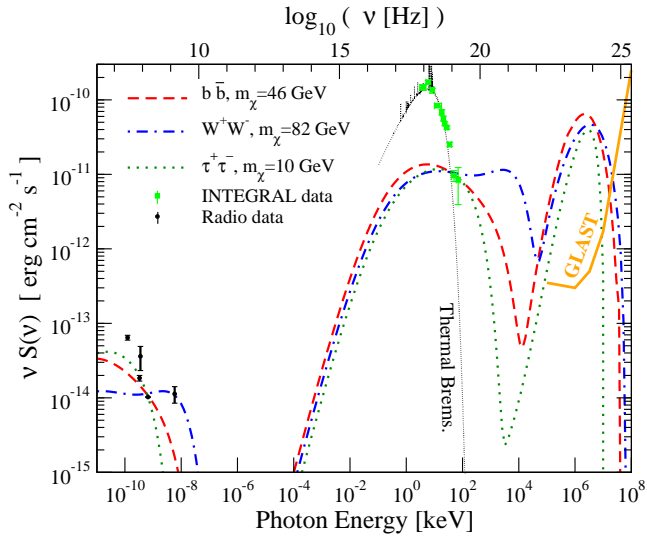}}
\caption{The spectral energy distribution for the multi-wavelength emission of three DM particle models and for the thermal X-ray emission. We also show the INTEGRAL \cite{Eckert:2007fv} and radio \cite{radiodata} data.}
\label{fig:SED}
\end{figure}
We illustrate the whole broad-band spectrum for the DM annihilation interpretation of the INTEGRAL hard X-ray emission in Fig.~\ref{fig:SED}. Particularly crucial here is the question of whether the radio data on the Ophiuchus cluster are compatible with our predictions. This question depends on the assumed value of the average cluster magnetic field $B$. In the Figure we use values of $B$ giving the best fit to the radio data, namely $B(f=b\bar b)=0.15\ \mu$G, $B(f=W^+W^-)=0.1\ \mu$G and $B(f=\tau^+\tau^-)=0.18\ \mu$G. Clearly, a linear superposition of the various final states can yield a remarkably good fit to the available radio data. We also indicate the anticipated GLAST sensitivity, which shows that a sizable flux of $\gamma$-rays is expected if the DM annihilation scenario is indeed correct.

\begin{figure}
\centerline{\epsfig{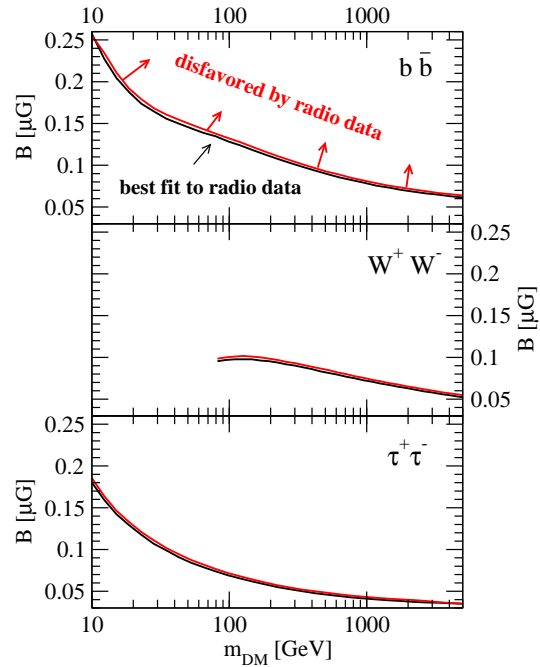}}
\caption{Preferred values of the cluster magnetic field as a function of the DM particle mass, for three representative pair annihilation final states.}
\label{fig:bmu}
\end{figure}
As pointed out in the analysis of Ref.~\cite{Eckert:2007fv}, simple IC interpretations of the INTEGRAL data force the estimate for the magnetic field in the range $B\sim 0.1-0.2\ \mu$G, which agrees with the present analysis. While this range is compatible with the value obtained from the hard X-ray emission in Coma \cite{FuscoFemiano:1999em}, it is below the estimates obtained, for other clusters, through Faraday rotation measures \cite{Kim} (see however the discussion in \cite{Pfrommer:2007sm}). We show in Fig.~\ref{fig:bmu} with black and red lines the values of $B$ respectively giving the best fit to radio data and exceeding by 5-$\sigma$ the measured radio flux in at least one bin. While values of $B$ above the red lines are disfavored by radio data, they are not strictly ruled out if one accounts, {\em e.g.}, for a radial dependence of the magnetic field \cite{Colafrancesco:2005ji}.

The detection of $\gamma$-rays from cosmic rays in clusters of galaxies is generically expected to be possible with GLAST \cite{Blasi:2007pm,Pfrommer:2007sm}. Specifically, the analysis of Ref.~\cite{andonagai} indicates that the Ophiuchus cluster will be detectable by GLAST provided the ratio of the energy density of cosmic rays to that of the thermal gas is larger than 0.3\% to 8.8\%, depending on assumptions on the cosmic ray spectral index and radial distribution. This warrants a systematic comparison between the expected $\gamma$-ray flux from cosmic rays and from DM in clusters, which is currently under way \cite{clustersinprep}. In general, though, the $\gamma$-ray spectra expected from cosmic rays (a power law above \mbox{$\sim\ 1$ GeV}, with a spectral index depending on the primary cosmic-ray protons spectral index) are significantly different from what expected from DM annihilation, see Fig.~\ref{fig:spectrum} and \ref{fig:SED}. If DM annihilation is responsible for most of the hard X-ray emission in the Ophiuchus cluster, GLAST will collect such large statistics that discrimination from a cosmic ray $\gamma$-ray emission appears reasonably feasible.

In summary, we showed that the origin of the non-thermal particles plausibly responsibly for the recently firmly discovered non-thermal hard X-ray emission from the Ophiuchus cluster might be generated by electrons and positrons produced in WIMP pair annihilations, provided the rate for the latter is large enough. This scenario is compatible with all observational information on the cluster, with particle DM production and searches, and, more importantly, will be thoroughly tested by GLAST. The future $\gamma$-ray telescope might even detect the monochromatic two-photon emission provided the particle model has a large enough branching ratio in that channel. Other galaxy clusters exhibiting non-thermal activity \cite{neva} will also be outstanding sites for GLAST to look for signatures of WIMP DM pair annihilation \cite{clustersinprep}. Finally, radio data from the Ophiuchus galaxy cluster can also be accounted for in the DM annihilation scenario, as long as the average magnetic field in the Ophiuchus cluster is below $\approx0.2\ \mu$G.

\begin{acknowledgments}
The author wishes to thank Tesla Jeltema for several insightful comments and for help with the computation of the thermal X-ray emission.
\end{acknowledgments}


\begin{thebibliography}{100}

\bibitem{Colafrancesco:2005ji}
  S.~Colafrancesco, S.~Profumo and P.~Ullio,
  Astron.\ Astrophys.\  {\bf 455} (2006) 21
  [arXiv:astro-ph/0507575].

\bibitem{WIMPReviews}
  G.~Jungman, M.~Kamionkowski, and K.~Griest,
  Phys.\ Rept.\  {\bf 267}, 195 (1996);
  G.~Bertone, D.~Hooper, and J.~Silk,
  Phys.\ Rept.\  {\bf 405}, 279 (2005).

\bibitem{KKDM}
  D.~Hooper and S.~Profumo,
  Phys.\ Rept.\  {\bf 453} (2007) 29
  [arXiv:hep-ph/0701197].

\bibitem{Colafrancesco:2006he}
  S.~Colafrancesco, S.~Profumo and P.~Ullio,
  Phys.\ Rev.\  D {\bf 75} (2007) 023513
  [arXiv:astro-ph/0607073].
  
\bibitem{xrgrinprep}
  T.~E.~Jeltema and S.~Profumo,
  ``{\em Searching for Dark Matter with X-ray Observations of Local Dwarf
  Galaxies}''
  arXiv:0805.1054 [astro-ph].

\bibitem{Eckert:2007fv}
  D.~Eckert, N.~Produit, S.~Paltani, A.~Neronov and T.~L.~Courvoisier,
  ``{\em INTEGRAL discovery of non-thermal hard X-ray emission from the Ophiuchus cluster}''
  arXiv:0712.2326 [astro-ph].
  

\bibitem{johnston}
M.~D.~Johnston {\em et al.}, Ap.J {\bf 245} (1981) 799.

\bibitem{radiodata}
B. Y. Mills, O. B. Slee, E. R. Hill,	
	Australian J. Phys., {\bf 13}, 676 (1960); 
O. B. Slee, Australian J. Phys., Ap. Suppl., {\bf 43}, 1 (1977);
O. B. Slee and C. S. Higgins, Australian J. Phys., Ap. Suppl., {\bf 36}, 60 (1975); J. R. Ehman, J. R. Dixon and J. D. Kraus, Ap. J. {\bf 75} (1970) 351; B. B. Jones and E. A. Finley,  Australian J. Phys., {\bf 27}, 687 (1974); 

\bibitem{Reimer:2003er}
  O.~Reimer, M.~Pohl, P.~Sreekumar and J.~R.~Mattox,
  Astrophys.\ J.\  {\bf 588} (2003) 155
  [arXiv:astro-ph/0301362].

\bibitem{feretti}
  L.~Feretti and G.~Giovannini,
  arXiv:astro-ph/0703494.

\bibitem{Rossetti:2003vy}
  M.~Rossetti and S.~Molendi,
  Astron.\ Astrophys.\  {\bf 414}, L41 (2004)
  [arXiv:astro-ph/0312447].


\bibitem{Pfrommer:2007sm}
  C.~Pfrommer,
  arXiv:0707.1693 [astro-ph].

\bibitem{FuscoFemiano:1999em}
  R.~Fusco-Femiano {\it et al.},
  Astrophys.\ J.\  {\bf 513}, L21 (1999)
  [arXiv:astro-ph/9901018].

\bibitem{Renaud:2006ch}
  M.~Renaud, G.~Belanger, J.~Paul, F.~Lebrun and R.~Terrier,
  arXiv:astro-ph/0606114.

\bibitem{FuscoFemiano:2005tz}
  R.~Fusco-Femiano, R.~Landi and M.~Orlandini,
  Astrophys.\ J.\  {\bf 624}, L69 (2005)
  [arXiv:astro-ph/0504147].


\bibitem{Sanders:2005jx}
   J.~S.~Sanders, A.~C.~Fabian, S.~W.~Allen and R.~W.~Schmidt,
  Mon.\ Not.\ Roy.\ Astron.\ Soc.\  {\bf 349} (2004) 952
  [arXiv:astro-ph/0311502]; J.~S.~Sanders, A.~C.~Fabian and R.~J.~H.~Dunn,
  Mon.\ Not.\ Roy.\ Astron.\ Soc.\  {\bf 360} (2005) 133
  [arXiv:astro-ph/0503318].
  
\bibitem{neva}
    J.~Nevalainen, T.~Oosterbroek, M.~Bonamente and S.~Colafrancesco,
  Astrophys.\ J.\  {\bf 608} (2004) 166
  [arXiv:astro-ph/0311142].

\bibitem{pythia}
 T.~Sjostrand, S.~Mrenna and P.~Skands,
  arXiv:0710.3820 [hep-ph].

\bibitem{kaastra}
J. S. Kaastra and R. Mewe, in Atomic Data Needs for X-ray Astronomy, p. 161, (2000), ed. by M. A. Bautista, T. R. Kallman, and A. K. Pradhan.

\bibitem{msugra}
  See {\em e.g.} H.~P.~Nilles,
  Phys.\ Rept.\  {\bf 110} (1984) 1.

\bibitem{focus}
  H.~Baer {\em et al.},
  JHEP {\bf 0510}, 020 (2005)
  [arXiv:hep-ph/0507282].

\bibitem{refsergio}
S.~Colafrancesco and P.~Blasi,
  Astropart.\ Phys.\  {\bf 9} (1998) 227
  [arXiv:astro-ph/9804262].


\bibitem{MayerHasselwander:1998hg}
  H.~A.~Mayer-Hasselwander {\it et al.},
  Astron.\ Astrophys.\  {\bf 335} (1998) 161.

\bibitem{Aharonian:2006wh}
  F.~Aharonian {\it et al.}  [H.E.S.S. Collaboration],
  Phys.\ Rev.\ Lett.\  {\bf 97}, 221102 (2006)
  [Erratum-ibid.\  {\bf 97}, 249901 (2006)]
  [arXiv:astro-ph/0610509].

\bibitem{N04}
 J.~F.~Navarro {\it et al.},
  Mon.\ Not.\ Roy.\ Astron.\ Soc.\  {\bf 349} (2004) 1039
  [arXiv:astro-ph/0311231].

\bibitem{bullock}
  J.~S.~Bullock {\it et al.},
  Mon.\ Not.\ Roy.\ Astron.\ Soc.\  {\bf 321}, 559 (2001)
  [arXiv:astro-ph/9908159].

\bibitem{reiprich}
T.~H.~Reiprich and H.~B\"oringer, Ap.J. {\bf 567} (2002) 716.

\bibitem{diemand}
  J.~Diemand, M.~Zemp, B.~Moore, J.~Stadel and M.~Carollo,
  Mon.\ Not.\ Roy.\ Astron.\ Soc.\  {\bf 364} (2005) 665
  [arXiv:astro-ph/0504215].

\bibitem{nonthermal}
See {\em e.g.} S.~Profumo and P.~Ullio,
  JCAP {\bf 0311} (2003) 006
  [arXiv:hep-ph/0309220] and R.~Catena {\em et al.}, arXiv:0712.3173 [hep-ph].

\bibitem{Cesarini:2003nr}
  A.~Cesarini, F.~Fucito, A.~Lionetto, A.~Morselli and P.~Ullio,
  Astropart.\ Phys.\  {\bf 21}, 267 (2004)
  [arXiv:astro-ph/0305075].

\bibitem{Profumo:2005xd}
  S.~Profumo,
  Phys.\ Rev.\  D {\bf 72}, 103521 (2005)
  [arXiv:astro-ph/0508628].

\bibitem{Evans:2003sc}
  N.~W.~Evans, F.~Ferrer and S.~Sarkar,
  Phys.\ Rev.\  D {\bf 69}, 123501 (2004)
  [arXiv:astro-ph/0311145].

\bibitem{Bergstrom:2005qk}
  L.~Bergstrom and D.~Hooper,
  Phys.\ Rev.\  D {\bf 73}, 063510 (2006)
  [arXiv:hep-ph/0512317].

\bibitem{Profumo:2006hs}
  S.~Profumo and M.~Kamionkowski,
  JCAP {\bf 0603}, 003 (2006)
  [arXiv:astro-ph/0601249].

\bibitem{Moskalenko:2006zy}
  I.~V.~Moskalenko, S.~W.~Digel, T.~A.~Porter, O.~Reimer and A.~W.~Strong,
  Nucl.\ Phys.\ Proc.\ Suppl.\  {\bf 173} (2007) 44
  [arXiv:astro-ph/0609768].

\bibitem{deBoer:2005tm}
  W.~de Boer, C.~Sander, V.~Zhukov, A.~V.~Gladyshev and D.~I.~Kazakov,
  Astron.\ Astrophys.\  {\bf 444} (2005) 51
  [arXiv:astro-ph/0508617].

\bibitem{Gebauer:2007mg}
  I.~Gebauer,
  arXiv:0710.4966 [astro-ph].

\bibitem{Kamionkowski:1994dp}
  M.~Kamionkowski, K.~Griest, G.~Jungman and B.~Sadoulet,
  Phys.\ Rev.\ Lett.\  {\bf 74} (1995) 5174
  [arXiv:hep-ph/9412213].






\bibitem{Profumo:2004at}
  S.~Profumo and C.~E.~Yaguna,
  Phys.\ Rev.\  D {\bf 70}, 095004 (2004)
  [arXiv:hep-ph/0407036].

\bibitem{latperform}
{\tt http://www-glast.slac.stanford.edu/}

\bibitem{ullioline}
L.~Bergstrom and P.~Ullio,
  Nucl.\ Phys.\  B {\bf 504} (1997) 27
  [arXiv:hep-ph/9706232].

\bibitem{Ferrer:2006hy}
  F.~Ferrer, L.~M.~Krauss and S.~Profumo,
  Phys.\ Rev.\  D {\bf 74} (2006) 115007
  [arXiv:hep-ph/0609257].

\bibitem{Kim}
K.-T.~Kim, P.~P. Kronberg and P.~C.~Tribble, Ap.J. {\bf 379} (1991) 80.

\bibitem{Blasi:2007pm}
  P.~Blasi, S.~Gabici and G.~Brunetti,
  Int.\ J.\ Mod.\ Phys.\  A {\bf 22} (2007) 681
  [arXiv:astro-ph/0701545].

\bibitem{andonagai}
S.~Ando and D.~Nagai,
  arXiv:0705.2588 [astro-ph].

\bibitem{clustersinprep}
J.~Kehayias, T.~Jeltema and S.~Profumo, in preparation.

\end{thebibliography}
\end{document}